# Theory of the ultrafast mode-locked GaN lasers in a large-signal regime

Igor V. Smetanin,[1] Peter P. Vasil'ev,[1]
and Dmitri L. Boiko[2*]

[1] P..N. Lebedev Physical Institute, 53 Leninsky prospect, Moscow 119991, Russia,
[2] CSEM, rue Jaquet-Droz 1, CH-2002 Neuchâtel, Switzerland.
[*] dmitri.boiko@csem.ch

**Abstract:** Analytical theory of the high-power passively mode-locked laser with a slow absorber is developed. In distinguishing from previous treatment, our model is valid at pulse energies well exceeding the saturation energy of absorber. This is achieved by solving the mode-locking master equation in the pulse energy-domain representation. The performances of monolithic sub-picosecond blue-violet GaN mode-locked diode laser in the high-power operation regime are analyzed using the developed approach.

**1. Introduction**

Compact laser sources generating sub-picosecond and femtosecond pulses in the blue/violet wavelength range [1] can meet the requirements for various applications in different fields of science and technology. They can be used for high-density optical data storage systems, biomedical diagnostic and high-resolution optical bio-imaging, optical combs, synchronization (time/frequency transfer) and entangled photon pair generation for quantum communications. For such an application, proper understanding of dynamic properties of GaN-based semiconductor lasers is particularly essential. Mode locking is often considered as the most promising technique for the generation of picosecond and femtosecond optical pulses with high repetition rates and low timing jitter [2,3].

One of the commonly used theoretical approaches to treat the mode-locking phenomenon in lasers with a slow gain and absorber has been elaborated by Haus and New in 70's [4,5]. Their approach utilizes a solitary time-domain master equation and considers localized gain and absorber. Due to its inherent simplicity, it has been proven to be very efficient. Several attempts have been made to adapt this efficient method to monolithic passively mode-locked laser diodes (LDs) [6,7]. However, all these previous developments assume low pulse energy as compared with the characteristic saturation energy of absorber. This assumption is not valid for the blue-violet GaN lasers: Using typical parameters of an InGaN/GaN laser diode, e.g. from Table 1, one obtains the saturation energy of absorber $E_{sA} = \hbar\omega v_g d h / \Gamma s (\partial g^*/\partial n)$ of 2.5 pJ for a single-mode ridge-waveguide structure. At the same time, the expected energy $W_p$ of the pulse travelling in the cavity is in the range 1-10pJ, which is comparable or higher then $E_{sA}$. This renders previous time-domain developments not suitable for monolithic mode-locked InGaN/GaN LDs.

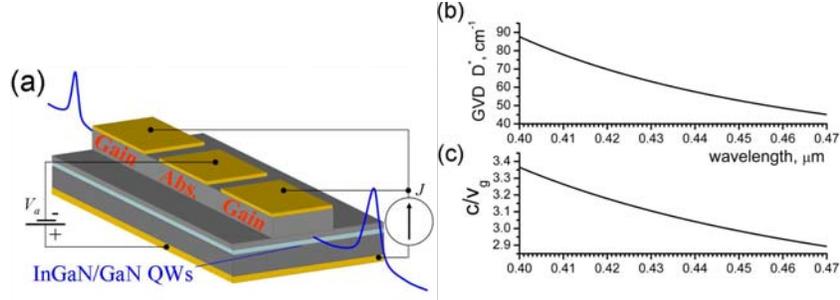

Fig. 1. Schematic of illustration of InGaN/GaN ridge-waveguide multi-section laser diode (a). Sellmeier model predictions for the group velocity dispersion $D^* = \omega_L^2 \partial^2 k / \partial \omega^2$ (b) and group velocity (c).

In this paper, we develop a new analytical approach for passively mode-locked diode lasers with slow absorber and amplifier. Our approach utilizes pulse-energy domain analysis and is suitable at high pulse energies, much exceeding the saturation energy of absorber section. We find that, two analytic solutions are possible, which are referenced herewith as a high- and low- energy branches. Ultrafast pulse operation mode-locking regime and its stability are discussed in application to InGaN/GaN monolithic multi-section laser diode.

**Table 1. Dynamic model parameters for monolithic mode-locked InGaN/GaN laser diodes, using data from Ref.[9].**

| Parameter | Value |
|---|---|
| $\hbar\omega$, photon energy | 2.9 eV |
| $\partial g^*/\partial n$, differential gain | $2.2 \times 10^{-6}$ cm$^3$/s |
| $s$, differential absorption to gain ratio | 3 |
| $n_t$ transparency carrier density | $1.6 \times 10^{19}$ cm$^{-3}$ |
| $h$ active layer thickness (QW width × QW number product) | 4 nm |
| $c/v_g$, group velocity index | 3.5 |
| $\Gamma$ mode confinement factor | 0.02 |
| $d$ waveguide width | 2 μm |
| $\alpha_i$ intrinsic material loss | 20 cm$^{-1}$ |
| $R_1 = R_2$ cavity facet reflection coefficients | 18.5% |
| $2\hbar\omega_L$ gain linewidth | 70 meV |
| $\alpha_{A,L}$ absorber (A) and amplifier (L) linewidth enhancement factors | 3 |
| $D^*$ dispersion coefficient | 3 |
| $T_L$ carrier lifetime in the gain section | 910 ps |
| $T_A$ absorber recovery time | 5 ps |
| $l_C$ cavity length | 0.8mm |

As a model system, we utilize monolithic multisection cavity, in which absorber section is located in the middle of the laser cavity (see Fig. 1.(a)). In fact, our approach and results applies well to any cavity arrangement, e.g. with absorber section located at the facet or in the external cavity configuration. The gain and absorption section are treated in localized dynamic model approximation.

## 2. Haus-New's theoretical model in time-domain representation

As a starting point, we use the steady-state master equation of the time-domain Haus-New's theory [4,5] which in the case of semiconductor laser diode can be written as [6,7]

$$\{1+(1-i\alpha_A)q_i \exp(-E(t)/E_{sA})-(1-i\alpha_L)g_i \exp(-E(t)/E_{sL})+\xi^2 +i(\psi+\xi)+ \\ +(1+\delta-2i\xi)\frac{1}{\omega_L}\frac{d}{dt}-(1+iD)\frac{1}{\omega_L^2}\frac{d^2}{dt^2}\}a(t)=0. \quad (1)$$

Here $a(t)$ is the complex slow varying amplitude (with respect to the carrier frequency) of the electric field component of the laser pulse, $E(t)=(dh/\Gamma)(e^2/8\pi c)\int_{-\infty}^{\infty}|a(t')|dt'$ and $W_p=(dh/\Gamma)(e^2/8\pi c)\int_{-\infty}^{\infty}|a(t')|dt'$ are the instantaneous (cumulative) and total pulse energies, $\xi=\Delta\omega/\omega_L$ is the frequency detuning of the carrier from the center of the gain line reported to the gain spectral half-width, $E_{sA}$ and $E_{sL}=sE_{sA}$ are the saturation energies of absorber and amplifier sections, $l_A$ and $l_L$ are their lengths. In Eq.(1), all material and geometrical parameters of the structure, as defined in Table 1, are normalized with respect to the cold cavity loss $\alpha_C l_C = \alpha_i l_C - (1/2)\ln(R_1 R_2)$ as indicated by the first term in Eq (1).

The carrier relaxation time $T_L$ in the gain-section quantum wells (QWs) and the recovery time $T_A$ of absorber are significantly longer than generated picosecond pulses (see Table 1). Therefore, large-signal absorption (second term in (1)) and gain (third term) are represented by instantaneous solutions of the corresponding rate equations during the build-up of an ultrafast pulse [6,7]. Here, $q_i$ and $g_i$ are the saturated absorption and gain in the cavity at the beginning of a pulse, $q_i\exp(-W_p/E_{sA})$ and $g_i\exp(-W_p/E_{sA})$ represent the same large-signal parameters at the end of the pulse. Solving the rate equations during the cavity roundtrip $T_{rep}$ time, when these relax back to the initial absorption $q_i$ and gain $g_i$, one obtains the relationships between the steady-state saturated and unsaturated small-signal values of the absorption and gain in the cavity [6,7]:

$$q_i = q_0 \frac{1-\exp(-T_{rep}/T_A)}{1-\exp(-W_p/E_{sA}-T_{rep}/T_A)}, \quad g_i = g_0 \frac{1-\exp(-T_{rep}/T_L)}{1-\exp(-W_p/E_{sL}-T_{rep}/T_L)} \quad (2)$$

where, for the model system in Fig.1, the normalized small-signal roundtrip absorption $q_0$ and gain $g_0$ in the cavity are controlled by the external bias $V_a$ and pump current density $J$:

$$q_0 = \frac{l_A \Gamma s}{v_g \alpha_C l_C}\left(\frac{\partial g}{\partial n}\right) n_t (1-V_a), \quad g_0 = \frac{l_L \Gamma}{v_g \alpha_C l_C}\left(\frac{\partial g}{\partial n}\right)\left[\frac{JT_L}{eh}-n_t\right]. \quad (3)$$

Here $n_t$ is the transparency carrier density, and $V_a$ is the normalized bias parameter.

The last four terms in Eq.(1) accounts for the dispersion coefficient $D^* = \omega_L^2 \partial^2 k/\partial\omega^2$, the round trip phase shift $\psi_0$ and time delay $\Delta T$ of the pulse. In Eq.(1), these are represented by normalized parameters $D=2D^*/\alpha_C$, $\psi=2\psi_0/\alpha_C l_C$, $\delta=2\omega_L \Delta T/\alpha_C l_C$ [6,7].

## 3. Analytic solution in the pulse-energy domain

So far the analytic solutions of the master Eq.(1) have been utilizing an assumption of small pulse energy $E(t), W_p \ll E_{sA,sL}$, yielding a series expansion of the gain and absorption terms in the time-domain [6,7]. We find an original approach to solve Eq.(1) for high pulse energies, developing our analysis in the pulse energy domain.

We introduce new variable $x = E(t)/W_p$, measuring the progress of instantaneous (cumulative) pulse energy towards the total energy in the pulse, so as $0 \leq x \leq 1$. It allows us to apply the following anzatz for the slowly varying pulse amplitude

$$a = \left[ AG^{1/2}(x) \right]^{1+i\beta}, \tag{4}$$

where $A$ is the peak pulse amplitude, $A = (4\pi c W_p / e^2 \tau_p)^{1/2}$, $\beta$ is the pulse chirp parameter, and the function $G(x)$ is the pulse intensity envelope. The function $G(x)$ reaches it's maximum value $\max(G) = 1$ at the peak of the laser pulse, and vanishes at the beginning and the end of the pulse, $G(0) = G(1) \equiv 0$. Substituting (4) in Eq.(1) and separating the real and imaginary parts, we transform it to the master equation for steady-state mode-locking regime in the energy domain representation:

$$\begin{aligned}
2(1-\beta D)B^2 GG'' &+ (1-\beta^2 - 2\beta D)B^2(G')^2 - (1+\delta + 2\xi\beta)BG' - \xi^2 = \\
&= 1 + q_i \exp(-\mu x) - g_i \exp(-\mu x/s), \\
2(\beta + D)B^2 GG'' &+ (2\beta + D - \beta^2 D)B^2(G')^2 - (\beta + \delta\beta - 2\xi)BG' = \\
&= \psi + \xi - \alpha_A q_i \exp(-\mu x) + \alpha_L g_i \exp(-\mu x/s)
\end{aligned} \tag{5}$$

where $\mu = W_p / E_{sA}$ and $B = 1/4\omega_L \tau_p$ are the normalized pulse energy and the inverse pulse width.

We obtain the analytic solution of Eq.(5) in the steady-state mode-locking regime by introducing a series expansion for the pulse envelop $G(x) \approx 1 - 4(x - 1/2)^2$ in the vicinity of the peak at $x=1/2$. Note that its inverse transform $t/\tau_p = \int_{1/2}^{x} dx'/G(x')$ to the time-domain representation yields the hyperbolic secant pulse shape $|a(t)|^2 = A^2 \cosh^{-2}(t/\tau_p)$. The term in the right-hand side of the first equation (5) is the net cavity gain $f(x) = -1 - q_i \exp(-\mu x) + g_i \exp(-\mu x/s)$. It reaches maximum at the pulse energy $x_{\max} = s \ln[sq_i / g_i]/\mu(s-1)$. Therefore, we substitute in Eq.(5) the series expansion $f(x) \approx s\Delta_1 - 1 - \frac{1}{2}\mu^2 \Delta_1 (x - x_{\max})^2$. In a similar way we introduce a series expansion in the right-hand side of the second equation (5). Finally, considering terms at each power of $x$, we obtain the following steady-state solution [Eqs. (6)-(10)], which is valid at any energy of the pulse, including the special case of our interest $\mu \equiv W_p / E_{sA} \geq 1$:

The chirp $\beta$ of the pulse (4) is

$$\beta = -\frac{2}{3} \frac{\gamma + D}{1 + D^2} Y. \tag{6}$$

where

$$\gamma = \left[ \frac{\alpha_L^s}{\alpha_A} \right]^{1/(s-1)}, \quad Y = \frac{3}{2} \frac{(1+D^2)}{(\gamma + D^2)} \left\{ \sqrt{2 + \frac{9}{4} \frac{(1-\gamma D)^2}{(\gamma + D)^2}} - \frac{3}{2} \frac{(1-\gamma D)}{(\gamma + D)} \right\}. \tag{7}$$

Using the notation

$$T_1 = Y^{-1} + 4Y\left(\frac{\beta+\gamma}{1+\beta^2}\right)^2 + \frac{2}{3}\frac{4(\gamma+D)-3(\gamma D-1)}{1+D^2}, \quad T_2 = 4b_1 + 8Y\left(\frac{\beta+\gamma}{1+\beta^2}\right)^2\frac{\beta b_1+\gamma b_2}{\beta+\gamma}$$

$$T_3 = 16\left(s-\frac{1}{\Delta_1}\right) - 8b_1^2 - 16Y\left(\frac{(\beta+\gamma)(\beta b_1+\gamma b_2)}{(1+\beta^2)(\beta+\gamma)}\right)^2, \tag{8}$$

$$\Delta_1 = \frac{s-1}{s}\left[\frac{(g_i/s)^s}{q_i}\right]^{1/(s-1)}, \quad b_1 = \frac{s}{s-1}\ln\left[\frac{sq_i}{g_i}\right], \quad b_2 = \frac{s}{s-1}\ln\left[\frac{\alpha_A}{\alpha_L}\frac{sq_i}{g_i}\right]$$

we can represent the normalized energy $\mu$ and duration $\tau_p$ of the mode-locked pulse in a simple form

$$\mu_\pm = \frac{T_2}{T_1} \pm \sqrt{\frac{T_2^2}{T_1^2}+\frac{T_3}{T_1}}, \quad \tau_{p,\pm} = \frac{4}{\omega_L}\frac{Y^{1/2}}{\mu_\pm\Delta_1^{1/2}}, \tag{9}$$

where the indexes "+" and "-" distinguish two solutions referenced here as the high-energy and low-energy branches, respectively; the FWHM of the hyperbolic secant pulse is $1.76\tau_p$. Finally, the normalized frequency detuning from the line center and the pulse delay read

$$\xi = \frac{1}{2}\frac{\beta+\gamma}{1+\beta^2}Y^{1/2}\Delta_1^{1/2}\left[\mu_\pm - 2\frac{\beta b_1+\gamma b_2}{\beta+\gamma}\right], \quad 1+\delta = \frac{1}{2}\frac{1-\beta\gamma}{1+\beta^2}Y^{1/2}\Delta_1^{1/2}\left[\mu_\pm - 2\frac{b_1-b_2\beta\gamma}{1-\beta\gamma}\right] \tag{10}$$

**4. Results and Discussion**

We apply our approach to analyze the stability and performances of a monolithic mode-locked InGaN/GaN diode laser. All major laser parameters are indicated in Table 1 and the ratio of the absorber length to the overall cavity length is $l_A/l_C = 0.1$. In Table 1, the short recovery time of absorber is settled by the tunneling of carriers from the QWs to the tilted QW barriers and by the carrier drift time through the intrinsic region of the multiple QW heterostructure barriers of about 30-40 nm thick. Thus, due to a large reverse bias, the absorber recovers much faster than the carrier relaxation takes place in the gain-section QWs. The group refractive index in InGaN/GaN lasers is close to 3.5 [8,9]. To estimate the group velocity dispersion we use Sellmaier model for the refraction index dispersion in GaN [8]. The predicted group velocity value (Fig.1.(c)) is in good agreement with the value from Table 1, therefore we are confident in the estimated group velocity dispersion of $D \approx 3$ in Fig. 1. (b).

The model predictions for the output peak power $P_\pm$ and FWHM pulse width ($1.76\tau_{p,\pm}$) are plotted in the Fig.2(a) as a function of the pump current. The output peak power per cavity facet is calculated as a fraction $\propto -\ln(R)/2(\alpha L - \ln R)$ of the pulse peak power in the cavity $W_p/2\tau_p$. In Fig. 2(b), the corresponding normalized pulse energy $\mu_\pm = W_{p,\pm}/E_{sA}$ and absorber bias parameter are displayed.

As expected, the higher the pulse energy, the shorter the pulse width is. This applies equally to the comparison of the pulse parameters between the high- (index "+") and low-energy (index "-") branches of Eq.(9). The relative energy $\mu_\pm$ of the mode-locked pulses traveling in the cavity is comparable to or greater than 1. Therefore our approach to solution of the Haus-New's master equation suitable for mode-locked pulses of arbitrary energy $W_p$ is not just an interesting feature of the model. As indicated in Fig 2(b), in InGaN/GaN monolithic mode-locked lasers, the pulse energy systematically exceeds the absorber saturation energy $E_{sA}$, and the absorber operates at high saturation level. Previously known solutions [6,7] are not valid in this operation regime of absorber. In the mode-locking regime corresponding to the high–energy branch of Eq.(9) the pulses of ~500fs width can be produced at the output peak pulse power of ~1.5W per laser facet.

The necessary conditions for stable regime of mode-locking has been formulated in Refs.[4,5,6,7]: The roundtrip net gain in the cavity must be negative all the time but the lasing pulse so as all spurious field fluctuations are dumped. In terms of the energy-domain solution (6)-(10), this requirement leads to the boundary conditions for $f(x)$:

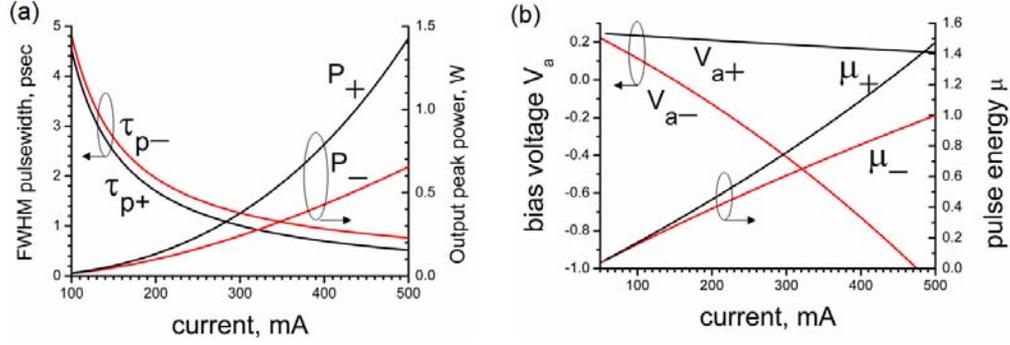

Figure 2. Solution Eq.(9) of the mode-locking master equation (1): (a) FWHM pulse width $1.76\tau_{p,\pm}$ (left axis) and output peak power per facet $P_\pm$ (righ axis) v.s. pumping current for high-energy (black curves) and low energy (red curves) solution branches; (b) Bias voltage at which the self-starting modelocking regime is stable (left axis) and the relative pulse energy $\mu_\pm = W_{p,\pm}/E_{sA}$ (righ axis)

$$1+q_i - g_i > 0 \quad (x=0), \quad 1+q_i \exp(-\mu_\pm) - g_i \exp(-\mu_\pm/s) > 0 \quad (x=1). \tag{11}$$

However during the pulse, the roundtrip net gain should become positive somewhere in the interval $0<x<1$ in the energy domain representation. Therefore the maximum of the roundtrip net gain $f(x)$ located at $x=x_{max}$ [see the discussion after Eq.(5)] assumes that $s\Delta_1 - 1 > 0$, yielding us the third condition for the steady-state mode-locking regime:

$$g_i \geq s q_i^{1/s} / (s-1)^{(s-1)/s} \tag{12}$$

It follows that there exists the minimum value of the absorption coefficient $q_{i\,min} = 1/(s-1)$ so as at $q_i < q_{i\,min}$ both the high-energy and low pulse energy mode-locking regimes are unstable. Using (11), (12) we calculate the domain of mode-locking stability is very narrow in the case of GaN laser diode and is shown in the absorber bias map in Fig.2(b) ..

We have to add also the necessary start-up condition

$$1+q_0 < g_0, \tag{13}$$

which means the net gain should be positive to launch the amplification from zero amplitude level.

## 5. Conclusion

In conclusion, we have developed a new analytical approach to the theory of passively mode-locked diode lasers at moderate pulse energies $\mu \geq 1$, which is based on the energy-domain representation of the pulse envelope evolution. In this approach, we calculate the parameters of the mode-locked pulse in the vicinity of the peak intensity at the arbitrary pulse energies, while usually [6,7] the small-energy approximation $\mu \ll 1$ is used which drops at pulse energies exceeding the absorber's saturation energy. The developed approximation allows us to determine the mode-locked operation parameters for the high-power sub-picosecond blue-violet GaN diode laser.

This research is supported by the EC Seventh Framework Programme under grant agreement #238556 (FEMTOBLUE)